\documentclass{elsart}

\usepackage{natbib}

\usepackage{amssymb}

\usepackage{amsmath}
\usepackage{mathrsfs}
\usepackage{graphicx}
\begin{document}

\begin{frontmatter}



\title{Effect of Internal Viscosity on Brownian Dynamics of DNA Molecules in Shear Flow}


\author[a,b]{Xiao-Dong Yang}
\author[b]{Roderick V. N. Melnik \corauthref{Cor}}

\address[a]{Department of Engineering Mechanics, Shenyang
Institute of Aeronautical Engineering, Shenyang 110136, China}
\address[b]{Mathematical Modelling \& Computational Sciences,Wilfrid
Laurier University, Waterloo,Ontario, Canada N2L 3C5}

\corauth[Cor]{Correspondence author: \ead{rmelnik@wlu.ca}}

\begin{abstract}
The results of Brownian dynamics simulations of a single DNA
molecule in shear flow are presented taking into account the effect
of internal viscosity. The dissipative mechanism of internal
viscosity is proved necessary in the research of DNA dynamics. A
stochastic model is derived on the basis of the balance equation for
forces acting on the chain. The Euler method is applied to the
solution of the model. The extensions of DNA molecules for different
Weissenberg numbers are analyzed. Comparison with the experimental
results available in the literature is carried out to estimate the
contribution of the effect of internal viscosity.
\end{abstract}


\end{frontmatter}



\section{Introduction}
Along with the progress in study of biological properties of polymer
(such as DNA) molecules, their mechanical features in solutions have
also been investigated recently\citep{Smith}. Modeling the
macromolecules is a key in studying the dynamics of single molecules
or solutions. The bead-spring chain model is a widely used model to
simulate the chain molecules. If the entropic springs connecting the
beads are assumed nonlinear, the governing equations of the system
can not be solved analytically. The development of efficient
numerical procedures becomes then necessary. Indeed, the increasing
numbers of beads in the molecule chain makes the numerical
simulation difficult and time-consuming. However, the dumbbell model
with only two beads is known to be computationally efficient.
Furthermore, it can yield good results if appropriate factors, such
as internal viscosity, hydrodynamic interactions, are considered.

There are two main approaches in investigating polymers in
solutions: one is to study the bulk rheology of polymeric solutions
\citep{Larson} and the other is to chase the dynamics of a single
molecules in detail \citep{Shaqfeh}. With the ability to examine
dynamics of macromolecule chains by fluorescence microscopy
measurement developed in recent years, the research on the
deformation and dynamics of single molecules have progressed
substantially.

In the study of bulk rheology of polymer solutions, the dynamics of
the flow of start-up or sudden cessation have been discussed
theoretically \citep{{Prabhakar},{Hua}} and experimentally
\citep{{SmithE},{Goff},{Goshen}}. \citet{Hua} studied the finitely
extensible nonlinear elastic (FENE) dumbbell model with internal
viscosity (IV). They discussed the contribution of IV to the bulk
rheology of polymer solutions.

On the other hand, by video fluorescence microcopy method,
\citet{Smith} experimentally studied the fractional extension of
$\lambda$-phage DNA molecules in steady shear flow for different
Weissenberg numbers that denote the dimensionless shear rates. They
also calculated the probability distribution of the extension by a
statistical method. Although the bulk rheology can be obtained by
further computation of the extension of the dumbbell model
molecules, the analysis of the extension and its distribution for
specific polymers analytically is still limited in the literature
\citep{Hur}.

In this paper, the FENE dumbbell model with IV proposed by
\citet{Hua} is employed to study the extensions of $\lambda$-phage
DNA molecules in steady shear flow. After numerical computation, the
results of molecule extension and probability distribution are
compared to that obtained in experiment by \citet{Smith}. The plot
of fractional extension for different Weissenberg numbers is also
superimposed by that of different models, e.g., the bead-spring
model with wormlike spring by \citet{Jendrejack} and classical FENE
dumbbell model by \citet{Hur}. We show that the FENE dumbbell model
with IV can yield good agreement in simulation of $\lambda$-phage
DNA molecules.

In Section 2, we give a brief description of the dumbbell model with
IV and its governing stochastic differential equation. In Section 3,
the details of the algorithm to solve the governing equation and the
values of parameters are given. The numerical results are presented
and discussed in Section 4, followed by concluding remarks.

\section{Model for the Dynamics of DNA molecules}

From classical physics viewpoint, the mass points, beads of the
dumbbell model are assumed to be subjected to three forces: 1) a
hydrodynamic force, which describes the resistance to motion of the
bead bathing in the solvent, 2) a connector force, which is denoted
by the spring between the beads, 3) a Brownian force, which is
caused by the rapidly bombarding of the beads by small solvent
molecules. The second force is due to the entropic force leading the
molecules to their statistical equilibrium positions. It is natural
to think that there exists, a forth force, the resistance to motion
experienced during changes in the molecule conformation. In what
follows, the internal viscosity force is introduced to reflect such
intramolecular resistance effect.
\begin{center}
\begin{tabular}{cc}
    \includegraphics[width=130pt]{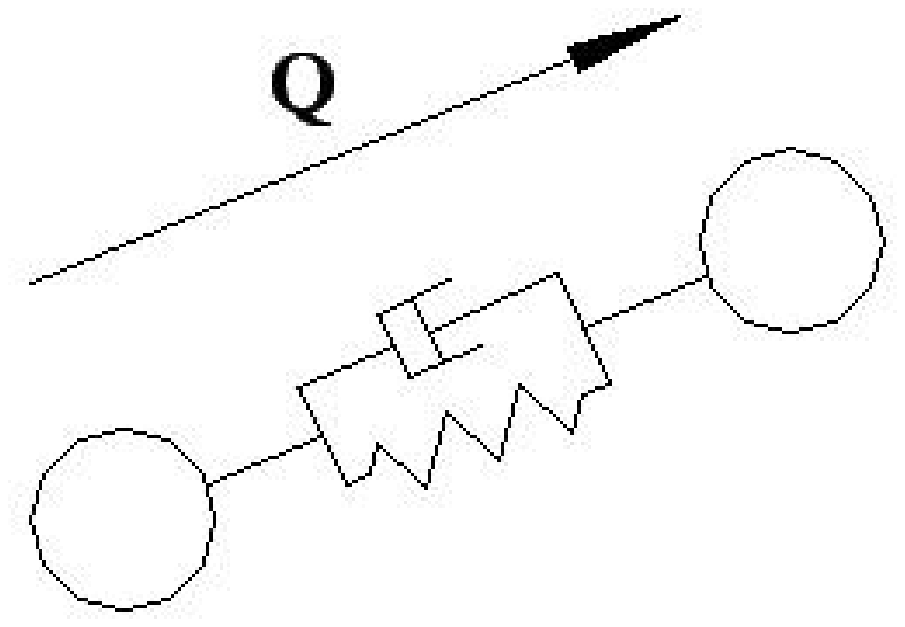} &
    \includegraphics[width=130pt]{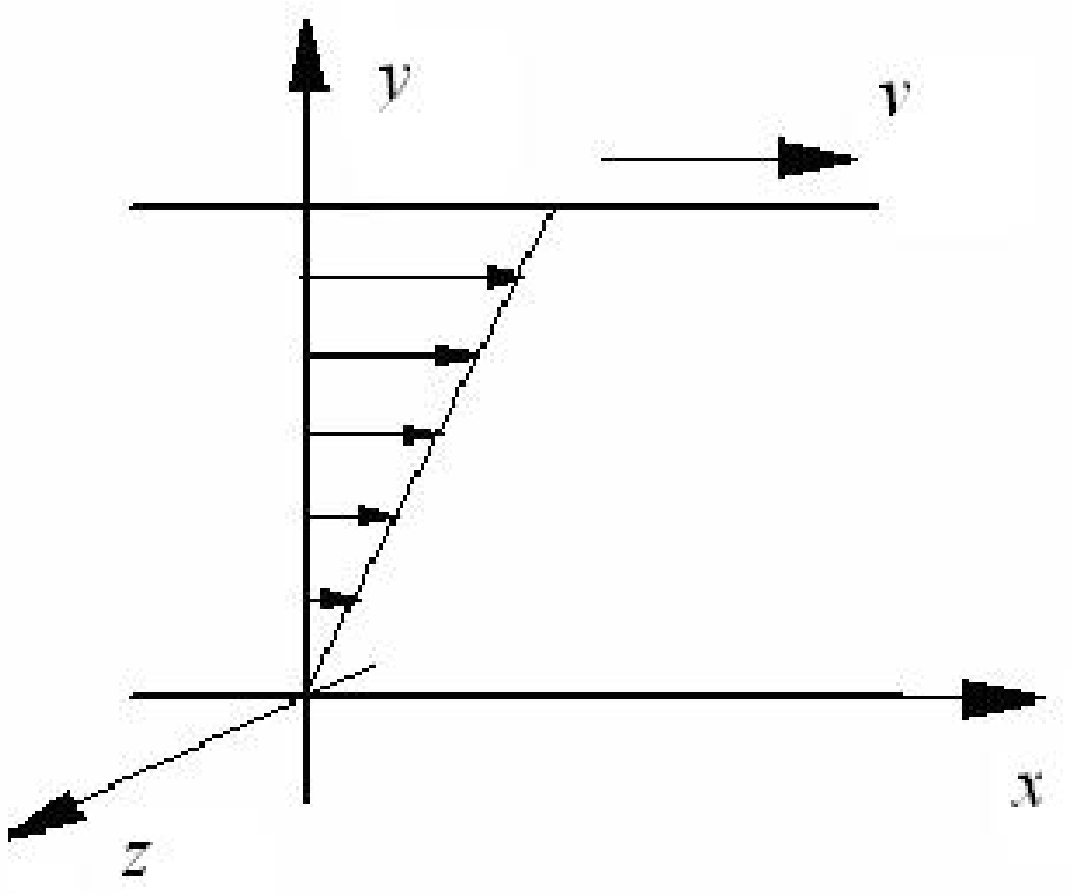}\\
    {\small Figure 1. The dumbbell model with
IV} & {\small Figure 2. The shear flow}
\end{tabular}
\end{center}
We start our consideration with the bead-spring-bead dumbbell model
of a single DNA molecule in a Newtonian solvent with viscosity $\eta
_s $. It is assumed that there is no interaction among the beads of
different dumbbells and the viscous drag coefficient is due to the
resistance of the flow denoted further by $\zeta $.

For the dumbbell model with internal viscosity, we use a schematic
representation of Figure 1 where $\bf{Q}$ is the connector vector of
the two beads. The spring force is a function of the configuration
vector and configuration velocity. This force is the combination of
connector force and the IV force mentioned in the first paragraph of
this section. If we consider the finitely extensible model, we get a
force law of the following form:
\begin{equation}
{\bf{F}}\left( {{\bf{Q}},\frac{{{\rm{d}}{\bf{Q}}}}{{{\rm{d}}t}}}
\right) = \frac{{H{\bf{Q}}}}{{1 - \left( {{Q \mathord{\left/
 {\vphantom {Q {Q_0 }}} \right.
 \kern-\nulldelimiterspace} {Q_0 }}} \right)^2 }} + K\left( {\frac{{{\bf{QQ}}}}{{Q^2 }}} \right)\frac{{{\rm{d}}{\bf{Q}}}}{{{\rm{d}}t}}
,
 \end{equation}
where extension $Q$ defined by $\sqrt{\bf{Q}\cdot\bf{Q}}$. In
equation (1), $Q_0$ is the maximum spring extension, and when the
extension of the dumbbell approaches this value, the spring force
tends to be infinite; $H$ and $K$ are Hookean constant and interval
viscous coefficient, respectively. The second moment tensor
$\bf{Q}\bf{Q}$ is a key to study the dynamics of the molecules. The
first term of the right side of equation (1) denotes the FENE spring
force proposed by \citet{Warner}. Considering the forces described
above, the equation of one bead's motion can be obtained if the
inertial term is neglected. According to the phase-space theorem
\citep{Bird}, the equation of motion can be given in the form of the
equation of the dumbbell configuration vector $\bf{Q}$:
\begin{equation}
{\frac{\rm{d}\bf{Q}}{\rm{d}t}} = \left( {{\boldsymbol{\delta }} -
\frac{1}{{\left( {{\zeta  \mathord{\left/
 {\vphantom {\zeta  {2K}}} \right.
 \kern-\nulldelimiterspace} {2K}}} \right) + 1}}\frac{{{\bf{QQ}}}}{{Q^2 }}} \right)
 \cdot \left( {\left[ {{\boldsymbol{\kappa }} \cdot {\bf{Q}}} \right] - \frac{{2kT}}{\zeta }
 \frac{\partial }{{\partial {\bf{Q}}}}\ln \boldsymbol \psi  - \frac{2}{\zeta }\frac{H}{{{1 - \left( {{Q \mathord{\left/
 {\vphantom {Q {Q_0 }}} \right.
 \kern-\nulldelimiterspace} {Q_0 }}} \right)^2 }}}{\bf{Q}}}
 \right),
 \end{equation}
where $\psi $
 is the configuration distribution function of $\bf{Q}$ changing
with time. The fluid velocity is given by specifying transpose of
imposed fluid velocity gradient ${\boldsymbol{\kappa
}}={\left(\boldsymbol{\nabla} \bf {v}\right)}^{\rm{T}}$ , and
$\boldsymbol\delta$ is unit matrix.

Now we introduce dimensionless parameters
 \begin{equation}
{\bf{\bar Q}} = \frac{{\bf{Q}}}{{\sqrt {{{kT} \mathord{\left/
 {\vphantom {{kT} H}} \right.
 \kern-\nulldelimiterspace} H}} }}
,\quad  \bar t = \frac{t}{\lambda } = \frac{{4Ht}}{\zeta },
\end{equation}
so that equation (2) could be cast into the dimensionless form
\begin{equation}
\frac{{{\rm{d}}{\bf{\bar Q}}}}{{{\rm{d}}\bar t}} = \left(
{{\boldsymbol{\delta }} - \frac{1}{{\left( {{\zeta  \mathord{\left/
 {\vphantom {\zeta  {2K}}} \right.
 \kern-\nulldelimiterspace} {2K}}} \right) + 1}}\frac{{{\bf{\bar Q\bar Q}}}}
 {{\bar Q^2 }}} \right) \cdot \left( {\left[ {\lambda {\boldsymbol{\kappa }} \cdot {\bf{\bar Q}}} \right]
 - \frac{1}{2}\frac{\partial }{{\partial {\bf{\bar Q}}}}\ln \psi  - \frac{1}{2}\frac{{{\bf{\bar Q}}}}{{1 - \left( {{{\bar Q} \mathord{\left/
 {\vphantom {{\bar Q} {\bar Q_0 }}} \right.
 \kern-\nulldelimiterspace} {\bar Q_0 }}} \right)^2 }}} \right),
 \end{equation}
where  $\lambda  = {\zeta \mathord{\left/
 {\vphantom {\zeta  {4H}}} \right.
 \kern-\nulldelimiterspace} {4H}}$ is the relaxation time of the molecules and
$\sqrt {{{kT} \mathord{\left/
 {\vphantom {{kT} H}} \right.
 \kern-\nulldelimiterspace} H}} $ is the root-mean-square average size of Hookean dumbbell in one dimension
at equilibrium. For convenience, the bars over the dimensionless
parameters are omitted thereafter without confusions.

Before we get the governing stochastic differential equation, the
Smoluchowski equation for this dumbbell model should be deduced,
which can be obtained by substituting equation (4) in equation of
continuity \citep{Bird}:
\begin{equation}
\begin{array}[50pt]{c}
\displaystyle\frac{{\partial \psi }}{{\partial t}} =  -
\frac{\partial }{{\partial {\bf{Q}}}} \cdot \left[ {\left(
{\boldsymbol{\delta}  - \frac{1}{{\varepsilon  +
1}}\frac{{{\bf{QQ}}}}{{Q^2 }}} \right) \cdot \left( {\lambda
{\boldsymbol{\kappa }} \cdot {\bf{Q}} -
\frac{1}{2}\frac{{\bf{Q}}}{{1 - \left( {{Q \mathord{\left/
 {\vphantom {Q {Q_0 }}} \right.
 \kern-\nulldelimiterspace} {Q_0 }}} \right)^2 }}} \right)\psi } \right]
 +\\ \noalign{\smallskip}
 \displaystyle\frac{1}{2}\frac{\partial }{{\partial {\bf{Q}}}}\left( {\boldsymbol{\delta}  - \frac{1}{{\varepsilon  + 1}}
 \frac{{{\bf{QQ}}}}{{Q^2 }}} \right) \cdot \frac{\partial }{{\partial
 {\bf{Q}}}}\psi,\\
 \end{array}
\end{equation}
where the internal viscosity parameter is defined as $\varepsilon  =
{{2K} \mathord{\left/
 {\vphantom {{2K} \zeta }} \right.
 \kern-\nulldelimiterspace} \zeta }$.

With the Ito interpretation we can obtain the equivalent stochastic
differential equation of equation (5) (see, e.g., \cite{Hua}):
\begin{equation}\begin{array}{c}
 \displaystyle {\rm{d}}{\bf{Q}} = \left[ {\left( {{\boldsymbol{\delta }} -
 \frac{\varepsilon }{{\varepsilon  + 1}}\frac{{{\bf{QQ}}}}{{Q^2 }}}
 \right) \cdot \left( {{\lambda \boldsymbol{\kappa }} \cdot {\bf{Q}} -
  \frac{1}{2} \frac{\bf{Q}}{1-\left( {{Q \mathord{\left/
 {\vphantom {Q {Q_0 }}} \right. \kern-\nulldelimiterspace} {Q_0 }}} \right)^2
}} \right) - \frac{\varepsilon }{{\varepsilon  +
1}}\frac{{\bf{Q}}}{{Q^2 }}} \right]{\rm{d}}t \\ \noalign{\smallskip}
 \displaystyle\quad {\kern 1pt} \quad \, + \left[ {{\boldsymbol{\delta }} - \left( {1 -
 \sqrt {\frac{1}{{\varepsilon  + 1}}} } \right)\frac{{{\bf{QQ}}}}{{Q^2 }}} \right] \cdot {\rm{d}}{\bf{W}}_t,  \\
 \end{array}
 \end{equation}
where
\begin{equation}\begin{array}{l}
 \displaystyle\left\langle {{\rm{d}}{\bf{W}}_t } \right\rangle  = {\bf{0}}, \\
 \displaystyle\left\langle {{\rm{d}}{\bf{W}}_t {\rm{d}}{\bf{W}}_t } \right\rangle  = \delta \left( {t - t'} \right){\bf{\delta }}{\rm{d}}t. \\
 \end{array}
 \end{equation}
If $\varepsilon =0$ in equation (6), the stochastic differential
equation degenerates to case for the classical FENE dumbbell case as
in the references by \cite{Hur} and \cite{Ottinger}.

In the next section, the algorithm and the specific parameter values
will be presented in detail for $\lambda$-phage DNA in steady shear
flow.
\section{Computational Implementation}

 The steady state simple shear flow in the Cartesian coordinate system
has the velocity vector $\bf v$ in three dimensions, with elements
$v_x=\dot{\gamma}y, v_y=0, v_z=0$ and where
$\dot{\gamma}={{{\rm{d}}v_x } \mathord{\left/
 {\vphantom {{{\rm{d}}v_x } {{\rm{d}}y}}} \right.
 \kern-\nulldelimiterspace} {{\rm{d}}y}}$ is the shear rate (See Figure 2). In such flow, $\boldsymbol\kappa$ can be expressed in linear function of shear rate
$\dot{\gamma}$
\begin{equation}
\boldsymbol \kappa  = \left( {\begin{array}{*{20}c}
   0 & \dot \gamma & 0  \\
   0 & 0 & 0  \\
   0 & 0 & 0  \\
\end{array}} \right).
\end{equation}
So we can adjust the parameter $\lambda\boldsymbol\kappa$ in
equation (6) by tuning Weissenberg number $W_i$ defined by
$W_i=\lambda\dot\gamma$.

Based on Euler's method, equation (6) can be discretized as:
\begin{equation}\begin{array}{c}
 \displaystyle{\bf{Q}}_{t+\Delta t} ={\bf{Q}}_{t}  \left[ {\left( {{\boldsymbol{\delta }} -
 \frac{\varepsilon }{{\varepsilon  + 1}}\frac{{{{\bf{Q}}}_{t}{\bf{Q}}_{t}}}{{Q_t^2 }}}
 \right) \cdot \left( {W_i \cdot {\bf{Q}}_{t} -
  \frac{1}{2} \frac{{\bf{Q}}_{t}}{1-\left( {{Q_{t} \mathord{\left/
 {\vphantom {Q {Q_0 }}} \right. \kern-\nulldelimiterspace} {Q_0 }}} \right)^2
}} \right) - \frac{\varepsilon }{{\varepsilon  +
1}}\frac{{\bf{Q}}_{t}}{{Q_{t}^2 }}} \right]\Delta t \\
\noalign{\smallskip}
 \displaystyle\quad {\kern 1pt} \quad \, + \left[ {{\boldsymbol{\delta }} - \left( {1 -
 \sqrt {\frac{1}{{\varepsilon  + 1}}} } \right)\frac{{{\bf{Q}}_{t}{\bf{Q}}_{t}}}{{Q_{t}^2 }}} \right] \cdot \Delta{\bf{W}}_t,  \\
 \end{array}
 \end{equation}
where $\Delta t$ is the constant time step and the components of
increments $\Delta {\bf{W}}_t={\bf{W}}_{t+\Delta t}-{\bf{W}}_t$,
which are all independent real-valued random variables with mean $0$
and variance $\Delta t$.

To solve equation (9) numerically, we need to give the values of
Weissenberg number, $W_i$, and the maximum spring extension, $Q_0$.
For convenience of comparison with the experimental results of
\citet{Smith}, Weissenberg number, decided by shear rate of the
solvent and relaxation time of DNA molecules, is chosen in the range
between $0$ to $80$. Further, the contour length of $\lambda$-phage
DNA is chosen as the extensibility parameter. So, the maximum spring
extension can be expressed as $Q_0=(N-1)a$, where $N$ and $a$ are
the number of beads and the length of a rod of the corresponding
bead-rod chain, respectively. All the parameters used in our
computation can be found in Table 1. In the computation, the initial
values of extension are chosen randomly as Gaussian distribution.

\begin{center}
Table 1. The value of parameters
\begin{tabular}{|c|c|c|c|}
\hline
Parameters & Values & Parameters & Values \\
\hline Weissengerg number $W_i$ & $0 \sim 80$ & No. bead-rod chain $N$ & $150$ \\
\hline Contour length of $\lambda$-DNA $L$ & $22\mu m$ & Length of rod $a$ & $0.147\mu m$ \\
\hline Time step $\Delta t$ & $10^{-5} \sim 10^{-3}$ & Internal viscosity $\varepsilon$ & $0.001$\\
\hline
\end{tabular}
\end{center}

For the FENE dumbbells, there is a certain probability that the
maximum allowed spring extension $Q_0$ is exceeded for any finite
time steps. To avoid this problem, we adopt the rejection method
that all moves larger than a fixed big value are rejected. Because
the extension that is very close to the maximum allowed value will
bring bad results, as proposed by \citet{Ottinger}, we reject all
moves which lead to a value of $Q^2$ larger than
\begin{equation}
\left( {1 - \sqrt {\Delta t} } \right)Q_0 ^2,
\end{equation}
and note that all the variables we use are in dimensionless form.
\begin{center}
    \includegraphics[width=250pt]{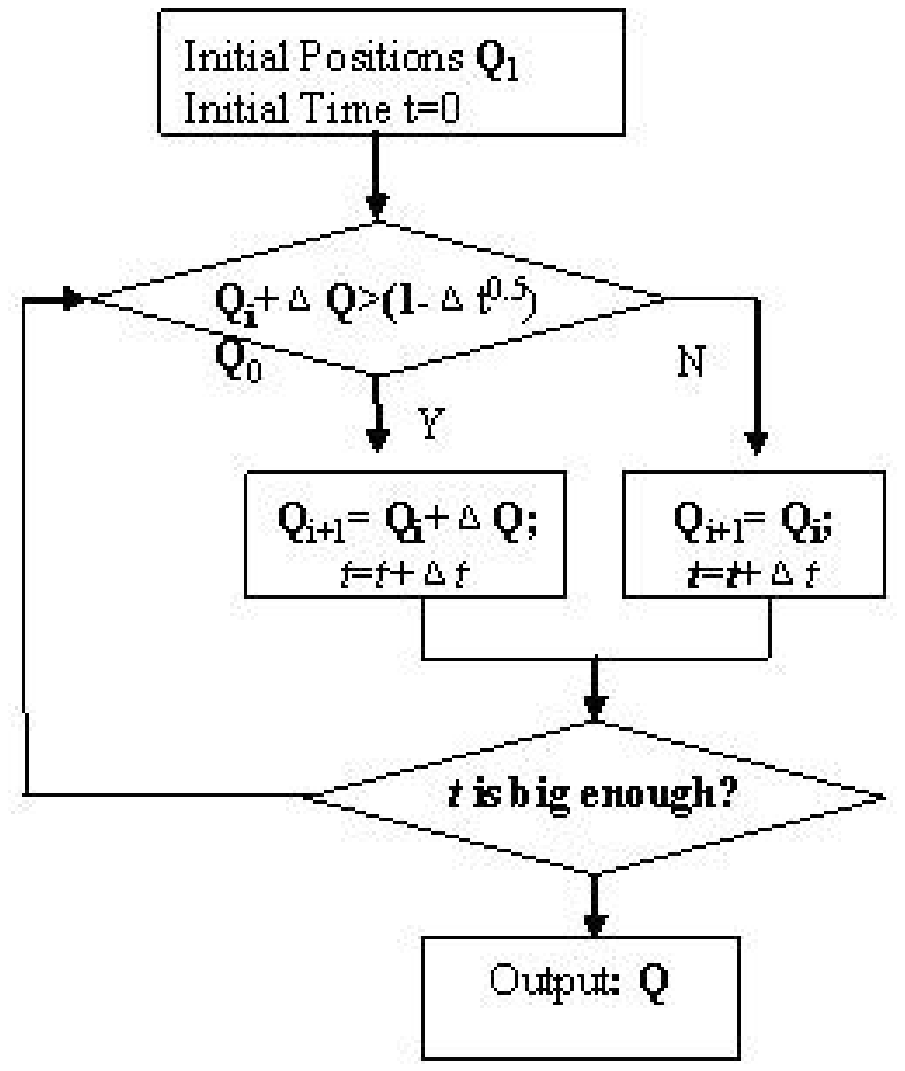}\\
\end{center}
\begin{center}{\small Figure 3. Flow chart of the simulation}\end{center}
After giving the initial position, the connector vector will be
calculated by the algorithm stated above. Figure 3 shows the flow
chart of the procedure. To obtain a good valuation of the mean
extension, $1000$ molecules or more should be considered, that is,
the procedure will be performed repeatedly to yield more accurate
mean value of the molecule extension.

\section{Numerical Experiments}

In this section, the computational method described in Section 3
will be applied to the solution of equation (9) to obtain the
extension distributions of the dumbbell of $\lambda$-phage DNA
moleculesfor different Weissenberg numbers. Simulation results will
be compared with the experimental data by \citet{Smith} and the
simulation results of FENE model \citep{Hur}, as well as with the
wormlike spring model with hydrodynamic interactions
\citep{Jendrejack}.

First, we study the extension's fluctuations along time. After the
numerical experiments with the parameters listed in Table 1 for
$2\times 10^6$ dimensionless time steps with $\Delta t=10^{-4}$, we
obtained the time history of one DNA molecule extension at
$W_i=6.3$, as demonstrated in Figure 4. The fractional extension
means the ratio between extension of one molecule with its contour
length, $Q/Q_0$. It can be found out from the raw data that there
are large fluctuations in the extension of the molecules even at low
shear rates.
\begin{center}
    \includegraphics[width=250pt]{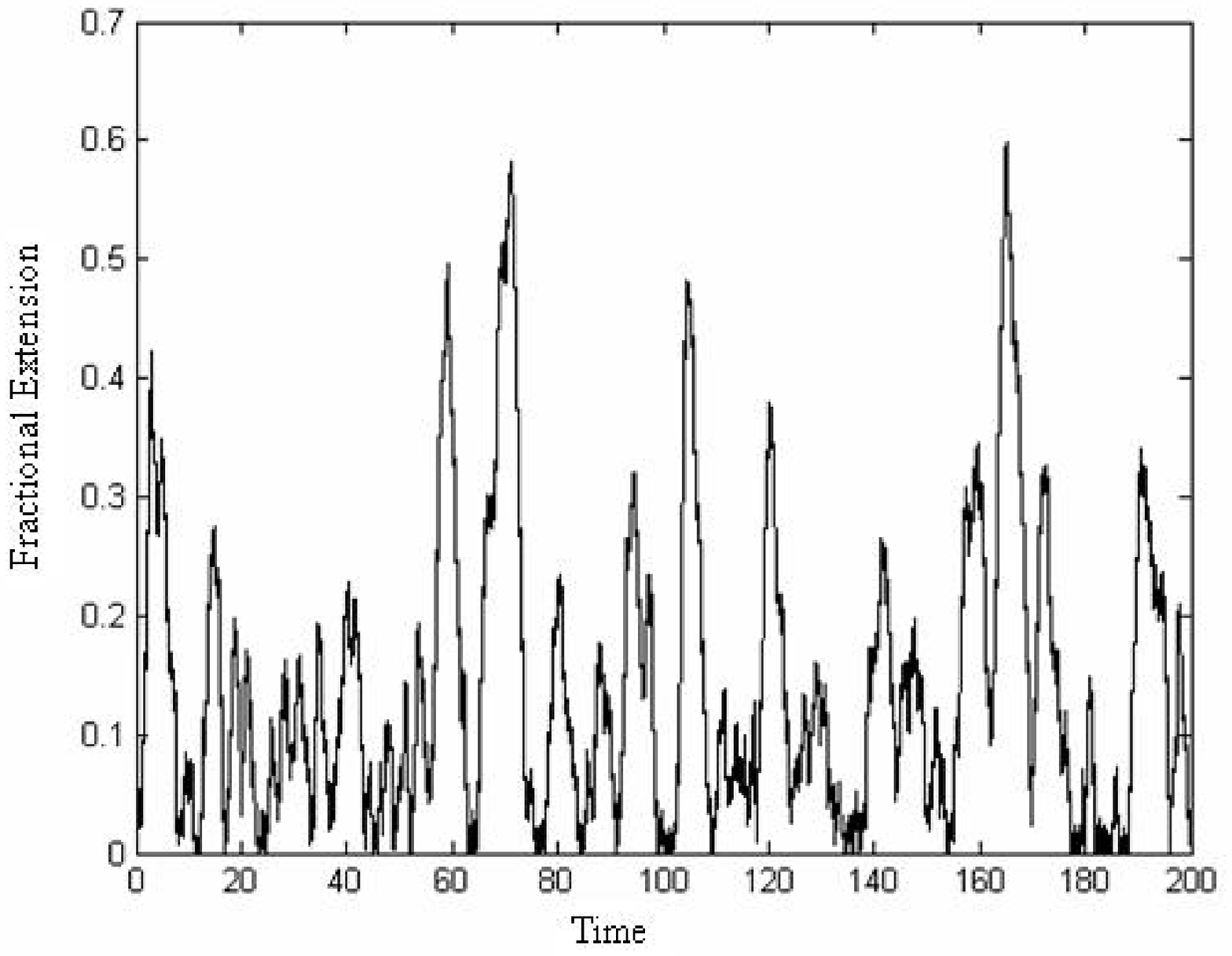}
\end{center}
\begin{center}{\small Figure 4. Dimensionless extension versus time}\end{center}
Next, we study the mean extension of 1000 molecules for different
Weissenberg numbers. In the simulation, the molecular extension is
projected in the $x$-$y$ plane to match Smith's experimental data
obtained by fluorescence microscopy with intensified video camera
perpendicular to the flow-vorticity plane. Every simulation with
different $W_i$ has given a specific mean extension value after
$10^6$ time steps when the average molecular extension appears
steady.

In Figure 5, we compare our simulation obtained with the FENE
dumbbell model accounting for internal viscosity with the
experimental data by \citet{Smith}. The dumbbell model without IV
\citep{Hur} and the wormlike model with hydrodynamic interactions
\citep{Jendrejack} are also superimposed in the plot. The classical
FENE dumbbell model can not predict the extension result when
Weissenburg number is high. Our dumbbell model and wormlike model
have given better values at high Weissenberg numbers.
\begin{center}
    \includegraphics[width=250pt]{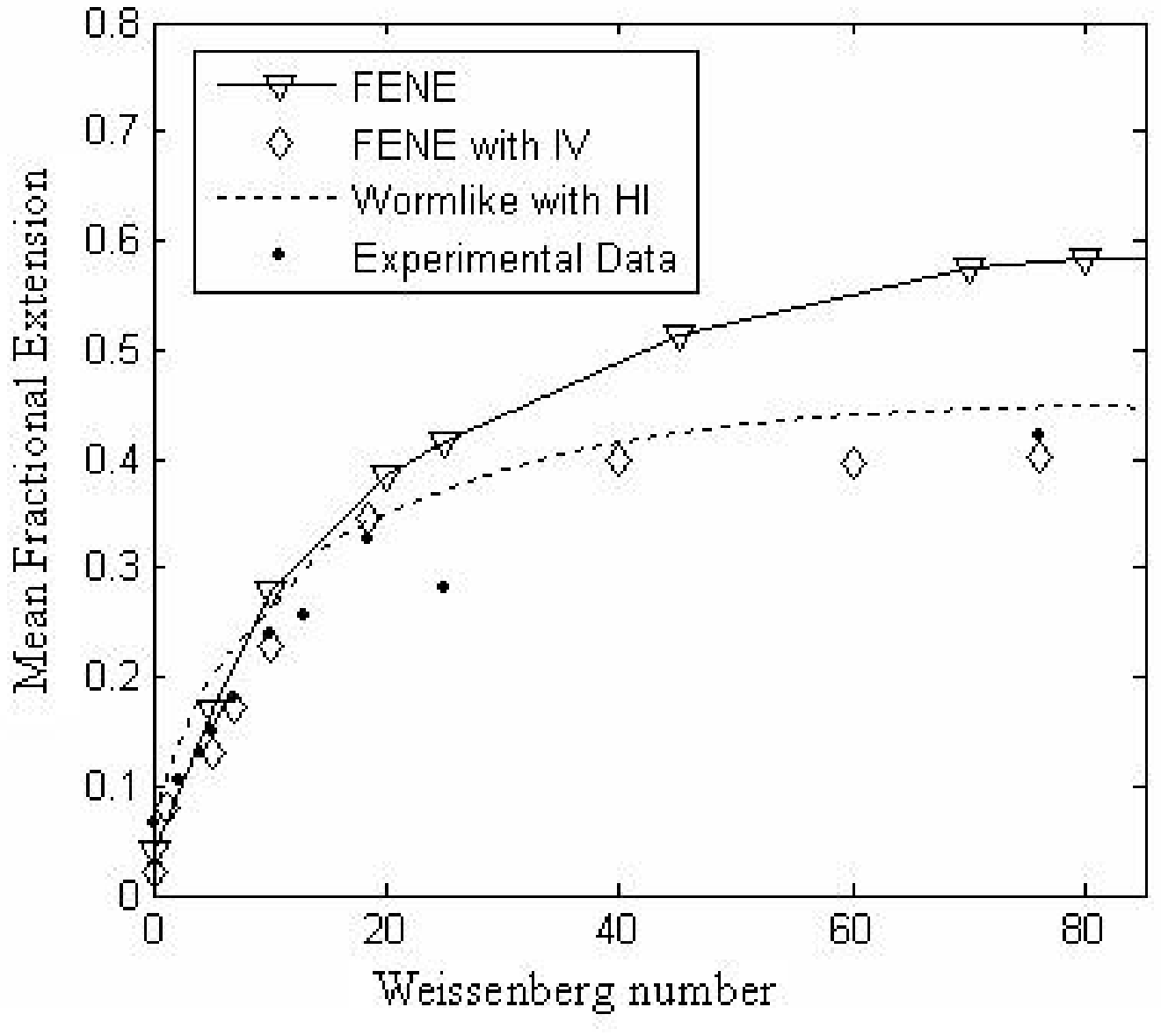}
\end{center}
\begin{center}{\small Figure 5. Dimensionless extension of DNA molecules versus $W_i$}\end{center}
When $W_i=0$, that is, the shear rate of the solvent vanishes, the
fluctuations of molecular extension are caused by the Brownian force
bombarding on DNA molecules of small solvent molecules. When $W_i$
increases, the extension reaches an asymptotic plateau as found by
\citet{Smith} and \citet{Hur}. Our simulation results have better
agreement with the experimental data compared to the other two
simulations presented in Figure 5, especially when the value of
Weissenberg number is high. We remark that this agreement with
available experimental data is obtained with a relatively simple
mathematical model where the internal viscosity effects were taken
into account.

Finally, we study the probability distribution of extension with the
numerical methodology described in Section 3. As shown in Figure 4,
from every snapshot of one DNA molecule we can get different values
of extension. Further, every molecule at the same time shows a us
different extension. Now, we investigate the probability
distribution of the extension of DNA molecules for some specific
Weissenberg numbers. After sampling 1000 DNA molecules during steady
state, we plotted the probability distribution function (PDF) of the
dimensionless extension in Figure 6. The PDF looks Gaussian in shape
at small Weissenberg number ($W_i=1.3$). Note further that when the
Weissenberg number increases, the peak of the PDF decreases and the
distribution broadens. At large Weissenberg numbers ($W_i=76$), the
PDF appears like white noise. For comparison, the experimental
results by \citet{Smith} were superimposed in Figure 6. As seen, the
simulation results have good qualitative and quantitative agreement
with the experimental results.

\begin{center}
    \includegraphics[width=290pt]{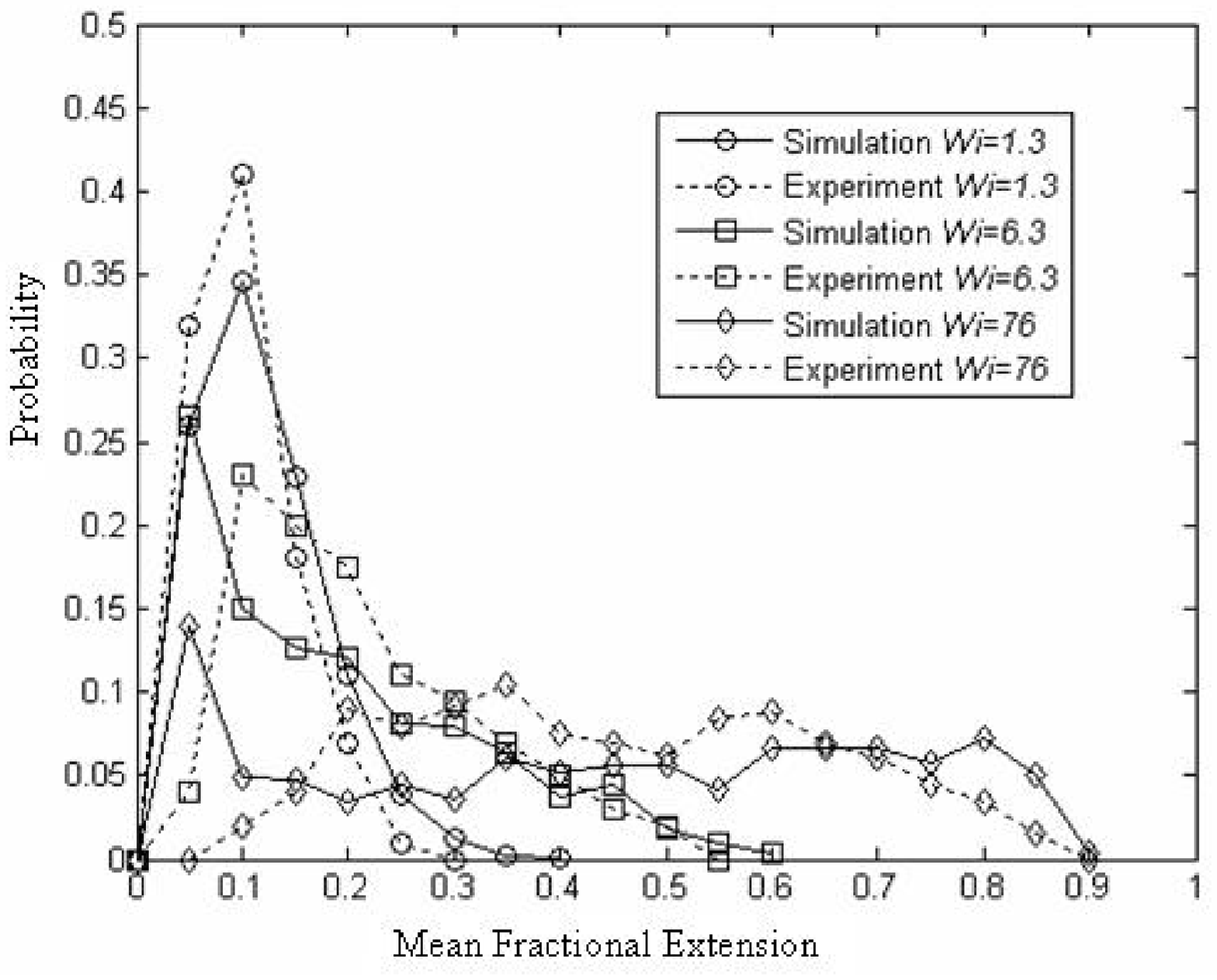}
\end{center}
\begin{center}
{\small Figure 6. The probability distribution function (PDF) of the
dimensionless extension}
\end{center}

The demonstrated agreement with the experimental results is
remarkable, given relative simplicity of the dumbbell model with IV
and its computational efficiency. The model and the developed
computational implementation provide a tool for the analysis of
extension and rheology properties.

\section{Conclusion}
In this paper, we focussed our analysis on the FENE dumbbell model
with internal viscosity for the single DNA molecule in shear flow.
The Brownian dynamics simulation was used to study the extension
between the two beads of the dumbbell. The results are in agreement
with the experimental data,  which is especially important for high
Weissenberg numbers. The probability distribution functions of the
extension were also presented and the qualitative and quantitative
agreement with available experiment data was demonstrated.

\end{document}